\documentclass[12pt]{article}

\usepackage{graphics}
\usepackage{epsfig}
\input epsf

\title{Deep inelastic prompt photon production at HERA \\ in the $k_T$-factorization approach}

\author{S.P.~Baranov$^a$, A.V.~Lipatov$^b$, N.P.~Zotov$^b$}

\begin{document}

\maketitle

\begin{center}

{\it $^a$\,P.N.~Lebedev Physics Institute,\\ 
119991 Moscow, Russia\/}\\[3mm]

{\it $^b$\,D.V.~Skobeltsyn Institute of Nuclear Physics,\\ 
M.V. Lomonosov Moscow State University,
\\119991 Moscow, Russia\/}\\[3mm]

\end{center}

\vspace{0.5cm}

\begin{center}

{\bf Abstract }

\end{center}

We investigate the prompt photon production in deep inelastic scattering at HERA 
in the framework of $k_T$-factorization QCD approach. Our study is
based on the off-shell partonic amplitude $e q^* \to e \gamma q$, where 
the photon radiation from the leptons as well as from the quarks is taken into account.
The unintegrated quark densities in a proton are determined using the Kimber-Martin-Ryskin prescription. 
The conservative error analisys is performed. 
We investigate both inclusive and jet associated prompt photon production rates.
Our predictions are 
compared with the recent
experimental data taken by the H1 and ZEUS collaborations.
We demonstrate that in the $k_T$-factorization approach 
the contribution from the quark radiation subprocess (QQ mechanism)
is enhanced as compared to the leading-order collinear approximation.

\vspace{0.8cm}

\noindent
PACS number(s): 12.38.-t, 13.85.-t

\vspace{0.5cm}

\section{Introduction} \indent 

The production of prompt photons\footnote{Usually the photons are called "prompt" 
if they originate from hard interaction, not from the decay of final-state hadrons.} 
in deep inelastic $ep$ collisions at HERA is a subject of 
intense studies~[1--3]. Such processes
provide a direct probe of the hard subprocess dynamics, since the produced photons 
are largely insensitive to the effects of final-state hadronization.
Studying the photon production in deep inelastic scattering (DIS) provides a test of
perturbative Quantum Chromodynamics (QCD) with two hard scales: the
transverse energy of the emitted photon $E_T^\gamma$, and the exchanged photon virtuality $Q^2$.
Good understanding of the production dynamics is also important for
searches of new particles decaying to photons at the LHC conditions.

At the leading-order, the production of prompt photons in deep inelastic $ep$ scattering
is described by the parton-level process $e q \to e \gamma q$, where
$q$ represents a quark or an anti-quark, and $e$ represents the beam lepton. The corresponding
cross section $\sigma(ep \to e\gamma X)$ can be obtained by convoluting the parton-level cross 
section $\hat \sigma(eq \to e \gamma q)$ with the quark distribution functions in the proton. 
The observed final-state photon can be emitted by a quark or by a lepton (the so-called 
QQ and LL mechanisms, respectively). The interference contribution (LQ)
is small. The QQ mechanism includes the direct photon radiation
from the quark line and also the so-called fragmentation processes~[4], where the produced quark 
forms a jet containing a photon with large fraction of the jet energy. 
This contribution involves poorly known quark-to-photon fragmentation 
functions $f_{q \to \gamma}(z)$~[5]. However, the isolation criterion
introduced in experimental analyses\footnote{See also Section~2.4.} substantially reduces the 
fragmentation component~[1--3].

Recently the H1 and ZEUS collaborations have reported their data~[1--3] on the 
deep inelastic production of prompt photons at HERA, both inclusive and 
in association with a hadronic jet. The presented data on the inclusive prompt photon production 
have been compared~[2, 3] with the LO pQCD ${\cal O}(\alpha^3)$ calculations~[6]. 
A substantial (by a factor of about 2) disagreement between the data and theory 
has been found~[3] at low $Q^2$.  
Also, the absolute size of the reported experimental cross sections turned out to
exceed the predictions of Monte-Carlo 
generators \textsc{herwig}~[7] and \textsc{pythia}~[8] (by the factors of 7.9 and 2.3,
respectively). Even after normalizing the total production rate, none of these programs
was able to describe all kinematical dependences of the measured cross sections~[1--3].
Comparison between the LO pQCD calculations~[6] and the data in bins of $\eta^\gamma$ shows that
the difference can be attributed~[3] mainly to the underestimation of 
the QQ contribution. At present, the next-to-leading order (NLO) ${\cal O}(\alpha^3 \alpha_s)$ predictions are only available~[9]
for the associated photon plus jet phase space selection, yet not for inclusive process. These predictions
are higher than the results of LO calculations~[6], especially at low $Q^2$, but still
underestimate the data~[3]. It was concluded~[1--3] that further theoretical investigations
(including the evaluation of higher-order processes) are needed to understand the 
observed discrepancy.

In the present paper to analyse the H1 and ZEUS data~[1--3] we use
instead the standard (collinear) approximation
the so-called $k_T$-factorization~[10] approach, which
is based on Balitsky-Fadin-Kuraev-Lipatov (BFKL) [11] or Ciafaloni-Catani-Fiorani-Marchesini 
(CCFM) [12] evolution equations.
A detailed description of the $k_T$-factorization approach can be
found, for example, in the reviews~[13]. 
Recently we have applied this approach to the inclusive and jet associated
prompt photon photoproduction at HERA~[14] and hadroproduction at Tevatron~[15]. 
As it was demonstrated in the ZEUS paper~[17] and also in the recent 
study~[17] performed
by the H1 collaboration, the $k_T$-factorization predictions are in better
agreement with the data than 
collinear NLO calculations. 

In the present study we extend our investigations~[14] to the DIS region. 
We calculate the relevant off-shell (i.e. $k_T$-dependent) amplitude ${\cal M}(e q^* \to e \gamma q)$,
where the virtuality of the incoming quarks is properly taken into account.
Our consideration covers 
both the LL and QQ production mechanisms. We neglect the LQ
mechanim since it gives only about 3\% contribution to the total 
cross section.
Then, we use the obtained expressions to calculate the total and differential cross sections of the
deep inelastic production of prompt photons (both inclusive and 
accompanied by a hadronic jet). 
The unintegrated quark densities in a proton $f_q(x,{\mathbf k}_{T}^2,\mu^2)$ are taken in
the Kimber-Martin-Ryskin (KMR) form~[18]. 
We make a systematic comparison of our predictions to the 
available H1 and ZEUS data~[1--3].
Our main goal is to study the capability of the $k_T$-factorization approach in
describing the DIS data. 
Such calculations are performed for the first time.

The outline of our paper is the following. In Section~2 we 
recall shortly the basic formulas of the $k_T$-factorization approach with a brief 
review of the calculation steps. 
In Section~3 we present the numerical results
of our calculations and a discussion. Section~4 contains our conclusions.
The compact analytic expressions for the off-shell matrix elements of   
the $e q^* \to e \gamma q$ subprocess are given in the Appendix. 

\section{Theoretical framework}
\subsection{Off-shell amplitude of the $e q^* \to e \gamma q$ subprocess} \indent 

Let us denote the four-momenta of the incoming electron
and proton and the outgoing electron, photon and final quark
by $p_e$, $p_p$, $p_{e^\prime}$, $p_\gamma$ and $p_{q}$, respectively. 
The initial quark and virtual photon have
four-momenta $k$ and $q$. 

Let us start with the LL production mechanism. There are two simple Feynman diagrams which describe this partonic
subprocess at the leading order in $\alpha_{\rm em}$.
The relevant matrix element reads
$$
  {\cal M}_{LL}(e q^* \to e \gamma q) = e_q e^3 \epsilon^\mu {1\over q^2} L_{\mu \nu} H^{\nu}, \eqno(1)
$$

\noindent
where $e_q$ is the fractional electric charge of the quark
and $\epsilon^\mu$ is the produced photon polarization four-vector.
The leptonic and hadronic tensors are
given by the following expressions:
$$
  L^{\mu \nu} = \bar u (p_{e^\prime}) \left[ \gamma^\mu \, {\hat p_{e^\prime} + \hat p_\gamma + m_e\over (p_{e^\prime} + p_\gamma)^2 - m_e^2}\, \gamma^\nu + \gamma^\nu \, {\hat p_e - \hat p^\gamma + m_e \over (p_e - p^\gamma)^2 - m_e^2} \, \gamma^\mu \right] u(p_e), \eqno(2)
$$
$$
  H^{\nu} = \bar u (p_q) \, \gamma^\nu \, u (k). \eqno(3)
$$

\noindent
When we calculate the matrix element squared, the summation over the produced photon polarizations is carried 
with $\sum \epsilon^\mu \epsilon^{  \, \nu} = - g^{\mu \nu}$,
and the spin density matrix for all on-shell spinors is taken in the standard 
form $u (p) \bar u (p) = \hat p + m$.
In the case of off-shell initial quarks the on-shell spin density matrix has to
be replaced with a more complicated expression. To evaluate it,
we "extend" the original diagram and consider the off-shell quark line as internal line in
the extended diagram. The "extended" process looks like follows:
the initial on-shell quark with four-momentum $p$ and mass $m_q$ radiates a quantum (say,
photon or gluon) and becomes an off-shell quark with four-momentum $k$.
So, for the extended diagram squared we write:
$$
  |{\cal M}|^2 \sim {\rm Sp} \left[ {\bar {\cal T}^\mu} \, {\hat k + m_q \over k^2 - m_q^2}\,\gamma^\nu \, u(p) \bar u (p) \, \gamma_\nu {\hat k + m_q \over k^2 - m_q^2} \, {\cal T}_{\mu} \right], \eqno(4)
$$

\noindent
where ${\cal T}$ is the rest of the original matrix element which remains unchanged.
The expression presented between $\bar {\cal T}^\mu$ and ${\cal T}_\mu$ now plays the role
of the off-shell quark spin density matrix. Using the on-shell condition 
$u(p) \, \bar u(p) = \hat p + m_q$
and performing the Dirac algebra one obtains in the massless limit $m_q \to 0$:
$$
  |{\cal M}|^2 \sim {1\over (k^2)^2} {\bar {\cal T}^\mu} \, \left( 2 k^2 \hat p - 4 (p \cdot k) \hat k \right) \, {\cal T}_{\mu}. \eqno(5)
$$

\noindent
Now we use the Sudakov decomposition $k = x p + k_T$ and neglect the second term 
in the parentheses in (5) in the small-$x$ limit to arrive at
$$
  |{\cal M}|^2 \sim {2\over x k^2} {\bar {\cal T}^\mu} \, x \hat p \, {\cal T}_{ \,\mu}. \eqno(6)
$$

\noindent
(Essentially, we have neglected here the negative light-cone momentum fraction of the 
incoming quark). The properly normalized off-shell spin density matrix is given by $x \hat p$,
while the factor $2/xk^2$ has to be attributed to the
quark distribution function (determining its
leading behavior). With this normalization, we successfully recover the on-shell collinear
limit when $k$ is collinear with $p$.

The evaluation of traces was done using the algebraic 
manipulation system \textsc{form}~[19]. The expression for 
$|\bar {\cal M}_{QQ}(e q^* \to e \gamma q)|^2$ can be easily obtained
from (1) --- (3) if we replace $p_e \to k$, $p_{e^\prime} \to p_q$, $k \to p_e$, $p_q \to p_{e^\prime}$
and multiply (1) by an extra factor $e_q$.
For the reader's convenience, we collect the 
analytic formulas for the off-shell matrix elements $|\bar {\cal M}_{LL}(e q^* \to e \gamma q)|^2$ 
and $|\bar {\cal M}_{QQ}(e q^* \to e \gamma q)|^2$ in the Appendix.

\subsection{Cross section for the prompt photon production} \indent 

According to the $k_T$-factorization theorem, the cross 
section of the process
$e p \to e \gamma X$ can be written as a convolution
of the off-shell matrix element $|\bar {\cal M}(e q^* \to e \gamma q)|^2$
and unintegrated quark distribution $f_q(x,{\mathbf k}_{T}^2,\mu^2)$:
$$
  \displaystyle \sigma_{LL,\,QQ}(e p \to e \gamma X) = \sum_q \int {1\over 256\pi^3 x^2 s \sqrt s |{\mathbf p}_{\gamma\, T}| \exp (y_\gamma)} |\bar {\cal M}_{LL,\,QQ}(e q^* \to e \gamma q)|^2 \, \times \atop 
  \displaystyle \times \, f_q(x,{\mathbf k}_{T}^2,\mu^2) d {\mathbf p}_{e^\prime \, T}^{2} d {\mathbf p}_{q \, T}^{2} d{\mathbf k}_{T}^2 d y_{e^\prime} d y_q \, {d\phi_{e^\prime} \over 2\pi} {d\phi_q\over 2\pi} {d\phi\over 2\pi}, \eqno(7)
$$

\noindent
where $y_{e^\prime}$, $y_{q}$ and $\phi_{e^\prime}$ and $\phi_{q}$ are the 
center-of-mass rapidities and azimuthal angles of the outgoing electron and (anti)quark, respectively.
The rapidity $y_\gamma$ of the produced photon is given by
$$
  y_\gamma = \ln \left[ { \sqrt s - m_{e^\prime \,T} \exp(y_{e^\prime}) - m_{q\,T} \exp(y_{q}) \over |{\mathbf p}_{\gamma\,T}| } \right], \eqno(8)
$$

\noindent
where $m_{e^\prime \,T}$ and $m_{q\,T}$ are the transverse masses of the corresponding particles.
We note that averaging the expression (7) over 
$\phi$ and taking the limit ${\mathbf k}_{T}^2 \to 0$ we
recover the well-known LO result of collinear parton model.
We use the unintegrated quark densities in a proton $f_q(x,{\mathbf k}_T^2,\mu^2)$ 
taken in the KMR form~[18].
The KMR approach is a formalism to construct the 
unintegrated parton (quark and gluon) distributions
from the known conventional parton
distributions $xa(x,\mu^2)$, where $a = g$ or $a = q$. 
For the input, we have used the recent leading-order Martin-Stirling-Thorne-Watt (MSTW)~[20]
parton densities.

As it is often done for 
deep inelastic prompt photon production, 
we choose the renormalization and factorization scales to be $\mu^2 = \xi Q^2$. 
In order to estimate the theoretical uncertainties of our calculations
we vary the scale parameter $\xi$ between 1/2 and 2 about the default value $\xi = 1$.
We use the LO formula for the strong coupling constant $\alpha_s(\mu^2)$ 
with $n_f = 4$ active (massless) quark flavours and $\Lambda_{\rm QCD} = 200$ MeV, 
such that $\alpha_s(M_Z^2) = 0.1232$. 

The multidimensional integration in (7) has been performed
by the means of Monte Carlo technique, using the routine \textsc{vegas}~[21].
The full C$++$ code is available from the authors on 
request\footnote{lipatov@theory.sinp.msu.ru}.

\subsection{Fragmentation contributions} \indent 

Perturbation theory becomes nonapplicable when the wavelength of the 
emitted photon (in the emitting quark rest frame) becomes larger then the 
typical hadronic scale ${\cal{O}}$ (1 GeV$^{-1}$). Then the nonperturbative 
effects of hadronization or fragmentation must be taken into account. 
Accordingly, the calculated cross section can be split into two pieces 
$$
  d\sigma = d\sigma_{\rm direct}(\mu^2) + d\sigma_{\rm fragm}(\mu^2) \eqno(9)
$$ 

\noindent
with $d\sigma_{\rm direct}(\mu^2)$ representing the perturbative 
contribution and $d\sigma_{\rm fragm}(\mu^2)$ the fragmentation 
contribution. In our calculations we choose the fragmentation scale 
$\mu^2$ to be the invariant mass of the quark + photon subsystem, 
$\mu^2 = (p + p_i)^2$, and restrict $d\sigma_{\rm direct}(\mu^2)$ 
to $\mu\ge M\simeq 1$~GeV. Under this condition, the contribution 
$d\sigma_{\rm direct}(\mu^2)$ is free from divergences. We have checked that the 
sensitivity of our results to the choice of $M$ is reasonably soft. 
As far as the fragmentation contribution is concerned, its 
size is dramatically reduced by the photon isolation cuts (see below). 
According to the estimates presented in~[22], the contribution 
from $d\sigma_{\rm fragm}$ amounts to only few percent of the visible cross 
section. This value is smaller than the theoretical uncertainty in 
the perturbative contribution $d\sigma_{\rm direct}$, and so, is 
neglected in our analysis.

\subsection{Photon isolation cuts} \indent 

In order to reduce huge background
from the secondary photons produced by the decays of $\pi^0$ and $\eta$ 
mesons the isolation criterion is introduced in the experimental analyses.
This criterion is the following. A photon is isolated if the 
amount of hadronic transverse energy $E_T^{\rm had}$, deposited inside
a cone with aperture $R$ centered around the photon direction in the 
pseudo-rapidity and azimuthal angle plane, is smaller than
some value $E_T^{\rm max}$:
$$
  \displaystyle \sum_{\rm had} E_T^{\rm had} \le E_T^{\rm max},\atop
  \displaystyle (\eta^{\rm had} - \eta)^2 + (\phi^{\rm had} - \phi)^2 \le R^2. \eqno(10)
$$

\noindent 
Both the H1 and ZEUS collaborations take $R \sim 1.0$ and 
$E_T^{\rm max} \sim 1$ GeV~[1--3]. 
Isolation not only reduces the background from 
light hadron decays
but also significantly reduces the fragmentation components
connected with collinear photon radiation. 
It was shown that after applying the isolation cut~(10) the 
contribution from the fragmentation subprocesses is suppressed~[22].

\section{Numerical results}

\subsection{Inclusive production} \indent 

Experimental data on the deep inelastic inclusive prompt photon production 
come from both the ZEUS and H1 collaborations. 
The differential cross sections are measured as function of 
the produced photon transverse energy $E_T^\gamma$, pseudo-rapidity $\eta^\gamma$, exchanged photon virtuality 
$Q^2$ and Bjorken variable $x$. 
The ZEUS data~[1] refer to the
kinematic region
defined as $5 < E_T^\gamma < 10$~GeV, $-0.7 < \eta^\gamma < 0.9$, $Q^2 > 35$~GeV$^2$,
$E_e^\prime > 10$~GeV, $139.8^{\rm o} < \theta_e^\prime < 171.8^{\rm o}$.
The initial electron and proton energies are $E_e = 27.6$ GeV and $E_p = 920$ GeV.
The $E_e^\prime$ and $\theta_e^\prime$ are the energy and the polar angle of the
scattered electron\footnote{All kinematic variables
are given in the laboratory frame where the positive direction of the OZ axis is given by the proton beam.}.
Another, very recent set~[2] of the ZEUS data, refers to the
kinematic region $4 < E_T^\gamma < 15$~GeV, $-0.7 < \eta^\gamma < 0.9$, $10 < Q^2 < 350$~GeV$^2$,
$E_e^\prime > 10$~GeV, $139.8^{\rm o} < \theta_e^\prime < 171.9^{\rm o}$. 
The invariant mass of the final state hadronic system $W_X > 5$~GeV.
The H1 data~[3] refer to the kinematic region
$3 < E_T^\gamma < 10$ GeV, $ - 1.2 < \eta^\gamma < 1.8$,
$4 < Q^2 < 150$~GeV$^2$, $E_e^\prime > 10$~GeV, $153 < \theta_e^\prime < 177$, $y > 0.05$. 
Additionally, the H1 data on the 
transverse energy distributions have been divided into five 
subdivisons of $\eta^\gamma$, namely $-1.2 < \eta^\gamma < -0.6$, 
$-0.6 < \eta^\gamma < 0.2$, $0.2 < \eta^\gamma < 0.9$, $0.9 < \eta^\gamma < 1.4$
and $1.4 < \eta^\gamma < 1.8$.
To suppress the contribution from the
elastic Compton scattering a cut on the mass of the final state hadronic system $W_X > 50$~GeV
is applied. 

Our predictions on the differential cross sections as a function of 
transverse energy $E_T^\gamma$, pseudo-rapidity $\eta^\gamma$, $Q^2$ and 
$x$ variables are shown in Figs.~1 --- 5 in 
comparison with the data~[1--3].
Solid histograms are obtained by fixing both the
factorization and normalization scales at the default value $\mu^2 = Q^2$,
whereas the upper and lower dashed histograms correspond to the scale variation as it
was described above.
We find that the predicted cross sections agree well with the H1 and ZEUS data
both in the normalization and shape. A slight underestimation of the H1 data
is only observed at high $Q^2$. In general, the $k_T$-factorization predictions
describe the data much better than the LO collinear calculations.

Our calculations give for the total cross sections $47.7_{-1.5}^{+1.3}$~pb, $6.1_{-0.3}^{+0.3}$~pb 
and $22.0_{-0.9}^{+0.7}$~pb for the H1~[3], ZEUS~[1] and ZEUS~[2] kinematic regions, respectively.
The experimentally measured cross sections are $50.3 \pm 1.7 \, {\rm (stat.)}\,^{+6.8}_{-7.8} \, {\rm (syst.)}$~pb,
$5.64 \pm 0.58 \, {\rm (stat.)}\,^{+0.47}_{-0.72} \, {\rm (syst.)}$~pb and
$19.4 \pm 0.7 \, {\rm (stat.)}\,^{+1.2}_{-1.0} \, {\rm (syst.)}$~pb.
The collinear LO pQCD predictions are 28.6~pb and 5.39~pb for the H1~[3] and ZEUS~[1] kinematical
regions. 
Besides that, the sole QQ contribution has been extracted in the ZEUS experiment~[2]
giving $12.2 \pm 0.7 \, {\rm (stat.)}\,^{+1.2}_{-1.0} \, {\rm (syst.)}$~pb. Our prediction 
of $14.8_{-0.5}^{+0.4}$~pb is very close to this value.
We find that the QQ contribution yelds about 65\% of the total cross section.
In the collinear LO calculations this mechanism gives only about 50\%. 
The reason for the enhancement of the QQ contribution in the $k_T$-factorization
approach can be seen in the fact that the unintegrated parton (quark) distribution
effectively include a large piece of higher-order corrections.
According to the observation of~[3], the difference  between the
collinear LO calculations and the data
can mainly be attributed to an underestimation of the QQ contribution.

The dependence of our results on the renormalization/factorization scale $\mu^2$ 
on its own weak, leading to about 5\% uncertainty band over a wide kinematic range. 
We cannot estimate the uncertainty connected with the choice of 
unintegrated quark distributions (although, it may be 
potentially much larger) because, at the time being, they are 
only available in the KMR scheme\footnote{Several attempts to calculate the CCFM-evolved unintegrated quark 
densities were made in~[23].} (see also reviews~[13] for more information).

\subsection{Production in association with a jet} \indent 

To calculate the jet-associated production of prompt photons 
we apply the procedure used previously in~[24].
The produced photon is
accompanied by a number of partons radiated in the course of the parton evolution.
On the average, the parton 
transverse momentum decreases from the hard interaction
block towards the proton. As an approximation, we assume that the parton with momentum $k'$ 
emitted at the last evolution step compensates the transverse momentum
of the parton participating in the hard subprocess, i.e. ${\mathbf k'}_{T} \simeq - {\mathbf k}_{T}$.
All the other emitted partons are collected together in the proton remnant, which
is assumed to carry only a negligible transverse momentum compared to ${\mathbf k'}_{T}$.
This parton gives rise to a final hadron jet with $E_T^{\rm jet} = |{\mathbf k'}_{T}|$,
in addition to the jet produced in the hard subprocess. From these hadron jets
we choose the one carrying the largest transverse energy.

Experimental data for this process have been obtained by the H1 and ZEUS
collaborations. The ZEUS collaboration presented the 
cross sections~[1] measured in the 
kinematic region $5 < E_T^\gamma < 10$ GeV, $E_T^{\rm jet} > 6$~GeV, 
$ - 0.7 < \eta^\gamma < 0.9$, $ - 1.5 < \eta^{\rm jet} < 1.8$,
$Q^2 > 35$~GeV$^2$, $E_e^\prime > 10$~GeV, $139.8^{\rm o} < \theta_e^\prime < 171.8^{\rm o}$
with electron energy $E_e = 27.6$ GeV and proton energy $E_p = 920$~GeV.
The more recent H1 data~[3] refer to the kinematic region
$3 < E_T^\gamma < 10$~GeV, $E_T^{\rm jet} > 2.5$~GeV, $ - 1.2 < \eta^\gamma < 1.8$,
$ - 1.0 < \eta^{\rm jet} < 2.1$, $4 < Q^2 < 150$~GeV$^2$, $E_e^\prime > 10$~GeV, $153^{\rm o} < \theta_e^\prime < 177^{\rm o}$,
$y > 0.05$ and $W_X > 50$~GeV with the same electron and proton energies.

The results of our calculations are shown in Figs.~6 and 7 in 
comparison with the HERA data. 
One can see that the distributions measured by the H1 collaboration
are well reproduced by our calculations. 
However, our results overeshoot the ealier ZEUS data.
A possible reason for that can be connected with the jet selection algorithm,
being the consequence of the used approximations.
Another possible reason 
is that our calculations are restricted to the parton level only,
while the jet fragmentation mechanism could also be of
importance.
Concerning the collinear NLO predictions, the shapes of the
differential cross sections $d\sigma/dE_T^\gamma$, $d\sigma/d\eta^\gamma$
and $d\sigma/dQ^2$ are described well, but the normalization
is too low, by about 40\%~[3]. The difference between the LO and NLO predictions
is mostly concentrated at low $Q^2$~[3].

The calculated total cross sections 
$\sigma(e p \to e \gamma + {\rm jet}) = 33.2_{-1.0}^{+0.9}$~pb and $1.9_{-0.1}^{+0.1}$~pb 
have to be compared with $31.6 \pm 1.2 \, {\rm (stat.)}\,^{+4.2}_{-4.8} \, {\rm (syst.)}$~pb and
$0.86 \pm 0.14 \, {\rm (stat.)}\,^{+0.44}_{-0.34} \, {\rm (syst.)}$~pb,
measured by the H1 and ZEUS collaborations in the relevant kinematic regions.   
We find perfect agreement between our predictions and recent H1 data.
A sizeble overestimation of the ZEUS data
is observed. It seems that the consistency of the H1 and ZEUS data
with each other can be questioned.
The total cross sections calculated in the
collinear approximation at the LO and NLO level for H1 conditions are 16.7~pb and 20.2~pb, respectively~[3].

As the final point, we should mention that corrections for hadronisation
and multiple interactions have been taken into account in the NLO analysis
of the HERA data~[1--3] performed in the framework of collinear factorization. 
The correction factors are typically 0.8 --- 1.2 depending on a bin.
These corrections are not taken into account in our consideration.

\section{Conclusions} \indent 

We have investigated the deep inelastic production of prompt photons at HERA in the
$k_T$-factorization approach. 
Our study is based on the off-shell partonic amplitude $e q^* \to e \gamma q$, where 
the photon radiation from the leptons as well as from the quarks is taken into account.
The unintegrated quark densities in a proton are calculated using the Kimber-Martin-Ryskin prescription. 

We have studied both the inclusive and jet associated production of prompt photons.
Our numerical predictions on the inclusive 
production cross sections are in well agreement with the H1 and ZEUS data. 
We have demonstrated that in the $k_T$-factorization approach 
the role of the QQ contribution 
is enhanced compared to the collinear LO approximation of QCD.
Our results for the jet associated 
production agree with the H1 measurements but overshoot the ZEUS data.
Probably, there exists a problem of incompatibility
of these H1 and ZEUS data with each other.

\section*{Acknowledgements} \indent 

The authors are very grateful to 
DESY Directorate for the support in the 
framework of Moscow --- DESY project on Monte-Carlo
implementation for HERA --- LHC.
A.V.L. was supported in part by the grants of the president of 
Russian Federation (MK-432.2008.2) and Helmholtz --- Russia
Joint Research Group.
Also this research was supported by the 
FASI of Russian Federation (grant NS-1456.2008.2),
FASI state contract 02.740.11.0244 and RFBR grant 08-02-00896-a.

\section {Appendix} \indent 

Here we present compact analytic expressions for the 
matrix elements involved in~(7). 
In the massless limit, the squared off-shell matrix elements $|\bar {\cal M}_{LL,\,QQ} (e q^* \to e \gamma q)|^2$ 
summed over the final states and 
averaged over the initial states are
$$
  |\bar {\cal M}_{LL}|^2 = {(4\pi \alpha_{em})^3 e_q^2 \, x \over 16 (k^2 - 2 (p_q \cdot k) )^2} \left[ {F_{LL}^{(1)} \over (p^\gamma \cdot p_{e^\prime})^2 } + {F_{LL}^{(2)} \over (p^\gamma \cdot p_e)^2} + { 2 F_{LL}^{(12)} \over (p^\gamma \cdot p_e) (p^\gamma \cdot p_{e^\prime}) } \right], \eqno(A.1)
$$

$$
  |\bar {\cal M}_{QQ}|^2 = {(4\pi \alpha_{em})^3 e_q^4 \, x \over 64 (p_e \cdot p_{e^\prime})^2} \left[ {F_{QQ}^{(1)} \over (p^\gamma \cdot p_q)^2 } + {F_{QQ}^{(2)} \over ( (p^\gamma \cdot k) - k^2)^2} + { 2 F_{QQ}^{(12)} \over (p^\gamma \cdot p_q) ( (p^\gamma \cdot k) - k^2 ) } \right], \eqno(A.2)
$$

\noindent
where
$$
  F_{LL}^{(1)} = 128 (p^\gamma \cdot p_q) (p^\gamma \cdot p_{e^\prime}) (p_e \cdot p_p) + 128 (p^\gamma \cdot p_{e^\prime}) (p^\gamma \cdot p_p) (p_q \cdot p_e), \eqno(A.3)
$$

$$
  F_{LL}^{(2)} = 128 (p^\gamma \cdot p_q) (p^\gamma \cdot p_e) (p_e \cdot p_p) + 128 (p^\gamma \cdot p_e) (p^\gamma \cdot p_p) (p_q \cdot p_{e^\prime}), \eqno(A.4)
$$

$$
  F_{LL}^{(12)} = - 64 (p^\gamma \cdot p_q) (p_e \cdot p_{e^\prime}) (p_e \cdot p_p) + 64 (p^\gamma \cdot p_q) (p_e \cdot p_{e^\prime}) (p_e \cdot p_p) - 
$$
$$
  64 (p^\gamma \cdot p_e) (p_q \cdot p_e) (p_e \cdot p_p) - 64 (p^\gamma \cdot p_e) (p_q \cdot p_{e^\prime}) (p_e \cdot p_p) - 
$$
$$
  128 (p^\gamma \cdot p_e) (p_q \cdot p_{e^\prime}) (p_e \cdot p_p) + 128 (p^\gamma \cdot p_{e^\prime}) (p_q \cdot p_e) (p_e \cdot p_p) + 
$$
$$
  64 (p^\gamma \cdot p_{e^\prime}) (p_q \cdot p_e) (p_e \cdot p_p) + 64 (p^\gamma \cdot p_{e^\prime}) (p_q \cdot p_{e^\prime}) (p_e \cdot p_p) - 
$$
$$ 
  64 (p^\gamma \cdot p_p) (p_q \cdot p_e) (p_e \cdot p_{e^\prime}) + 64 (p^\gamma \cdot p_p) (p_q \cdot p_{e^\prime}) (p_e \cdot p_{e^\prime}) - 
$$
$$
  128 (p_q \cdot p_e) (p_e \cdot p_{e^\prime}) (p_e \cdot p_p) - 128 (p_q \cdot p_{e^\prime}) (p_e \cdot p_{e^\prime}) (p_e \cdot p_p), \eqno(A.5)
$$

$$
  F_{QQ}^{(1)} = 128 (p^\gamma \cdot p_q) (p^\gamma \cdot p_e) (p_{e^\prime} \cdot p_p) + 128 (p^\gamma \cdot p_q) (p^\gamma \cdot p_{e^\prime}) (p_e \cdot p_p), \eqno(A.6)
$$

$$
  F_{QQ}^{(2)} =  128 (p^\gamma \cdot p_e) (p^\gamma \cdot p_p) (p_q \cdot p_{e^\prime}) + 128 (p^\gamma \cdot p_{e^\prime}) (p^\gamma \cdot p_p) (p_q \cdot p_e) + 
$$
$$ 
  128 (p^\gamma \cdot k) (p_q \cdot p_e) (p_{e^\prime} \cdot p_p) + 128 (p^\gamma \cdot k) (p_q \cdot p_{e^\prime}) (p_e \cdot p_p) - 
$$
$$
  128 (p^\gamma \cdot p_p) (p_q \cdot p_e) (p_{e^\prime} \cdot k) - 128 (p^\gamma \cdot p_p) (p_q \cdot p_{e^\prime}) (p_e \cdot k) -
$$
$$
  64 (p_q \cdot p_e) (p_{e^\prime} \cdot p_p) k^2 - 64 (p_q \cdot p_{e^\prime}) (p_e \cdot p_p) k^2, \eqno(A.7)
$$

$$
  F_{QQ}^{(12)} =  64 (p^\gamma \cdot p_q) (p_q \cdot p_e) (p_{e^\prime} \cdot p_p) + 64 (p^\gamma \cdot p_q) (p_q \cdot p_{e^\prime}) (p_e \cdot p_p) +
$$
$$
  64 (p^\gamma \cdot p_q) (p_e \cdot k) (p_{e^\prime} \cdot p_p) + 64 (p^\gamma \cdot p_q) (p_e \cdot p_p) (p_{e^\prime} \cdot k) + 
$$
$$
  64 (p^\gamma \cdot p_e) (p_q \cdot p_{e^\prime}) (p_q \cdot p_p) - 64 (p^\gamma \cdot p_e) (p_q \cdot k) (p_{e^\prime} \cdot p_p) +
$$
$$
  64 (p^\gamma \cdot p_{e^\prime}) (p_q \cdot p_e) (p_q \cdot p_p) - 64 (p^\gamma \cdot p_{e^\prime}) (p_q \cdot k) (p_e \cdot p_p) - 
$$
$$
  64 (p^\gamma \cdot k) (p_q \cdot p_p) (p_e \cdot p_{e^\prime}) - 128 (p^\gamma \cdot p_p) (p_q \cdot p_e) (p_q \cdot p_{e^\prime}) - 
$$
$$
  64 (p^\gamma \cdot p_p) (p_q \cdot p_e) (p_{e^\prime} \cdot k) - 64 (p^\gamma \cdot p_p) (p_q \cdot p_{e^\prime}) (p_e \cdot k) + 
$$
$$
  64 (p^\gamma \cdot p_p) (p_q \cdot k) (p_e \cdot p_{e^\prime}) - 64 (p_q \cdot p_e) (p_q \cdot k) (p_{e^\prime} \cdot p_p) - 
$$
$$
  64 (p_q \cdot p_e) (p_q \cdot p_p) (p_{e^\prime} \cdot k) - 64 (p_q \cdot p_{e^\prime}) (p_q \cdot k) (p_e \cdot p_p) - 
$$
$$
  64 (p_q \cdot p_{e^\prime}) (p_q \cdot p_p) (p_e \cdot k). \eqno(A.8)
$$

\newpage

\begin{figure}
\begin{center}
\epsfig{figure=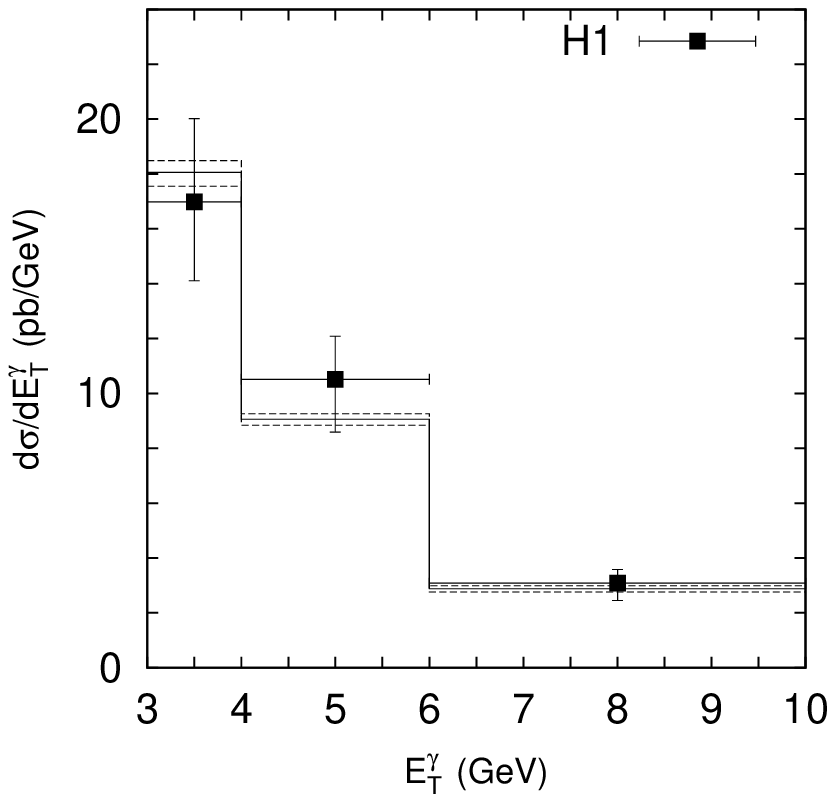, width = 8.1cm}
\epsfig{figure=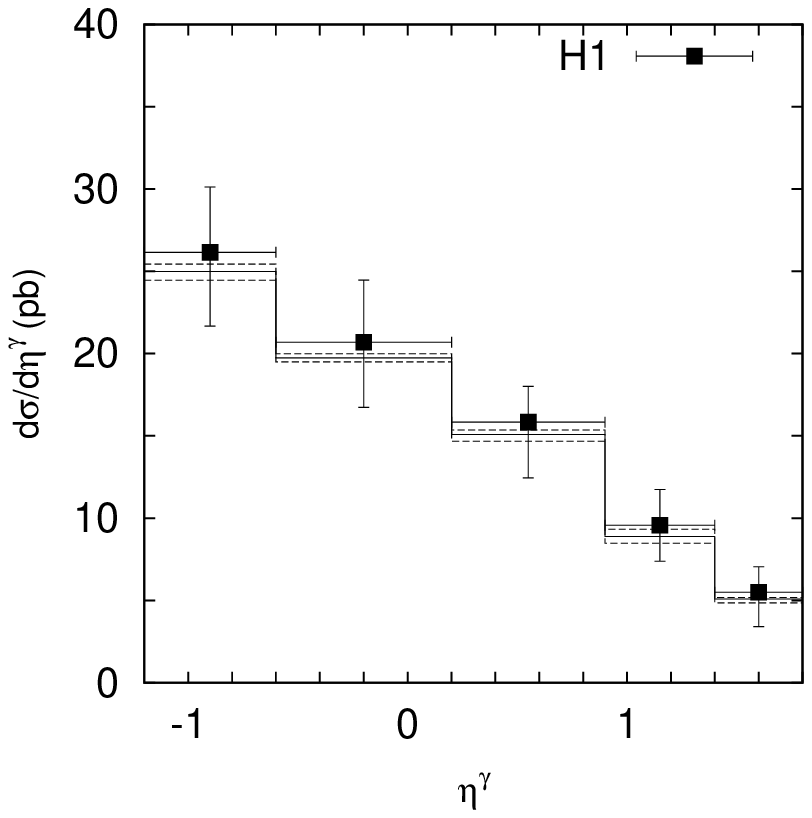, width = 8.1cm}
\epsfig{figure=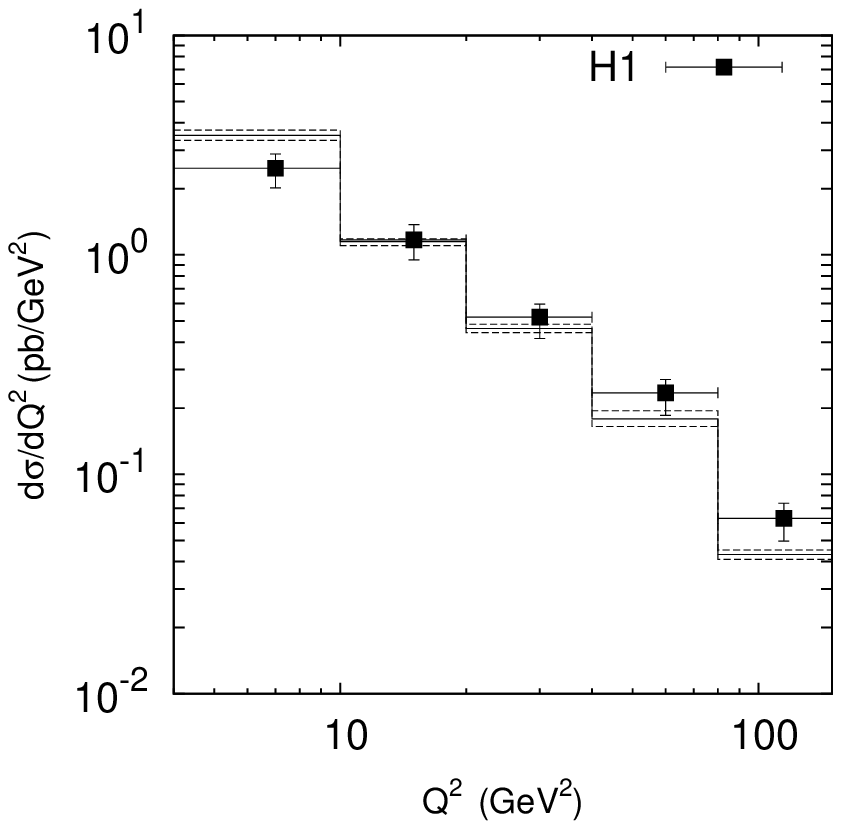, width = 8.1cm}
\caption{Differential cross sections of the 
inclusive deep inelastic prompt photon production
as a function of $E_T^\gamma$, $\eta^\gamma$ and $Q^2$ calculated
at $3 < E_T^\gamma < 10$ GeV, $ - 1.2 < \eta^\gamma < 1.8$,
$4 < Q^2 < 150$~GeV$^2$, $E_e^\prime > 10$~GeV, $153^{\rm o} < \theta_e^\prime < 177^{\rm o}$, $y > 0.05$
and $W_X > 50$~GeV.
The solid histogram corresponds to the default scale $\mu^2 = Q^2$, whereas the upper and 
lower dashed histograms correspond to scale variations described in the text.
The experimental data are from H1~[3].}
\end{center}
\label{fig1}
\end{figure}

\begin{figure}
\begin{center}
\epsfig{figure=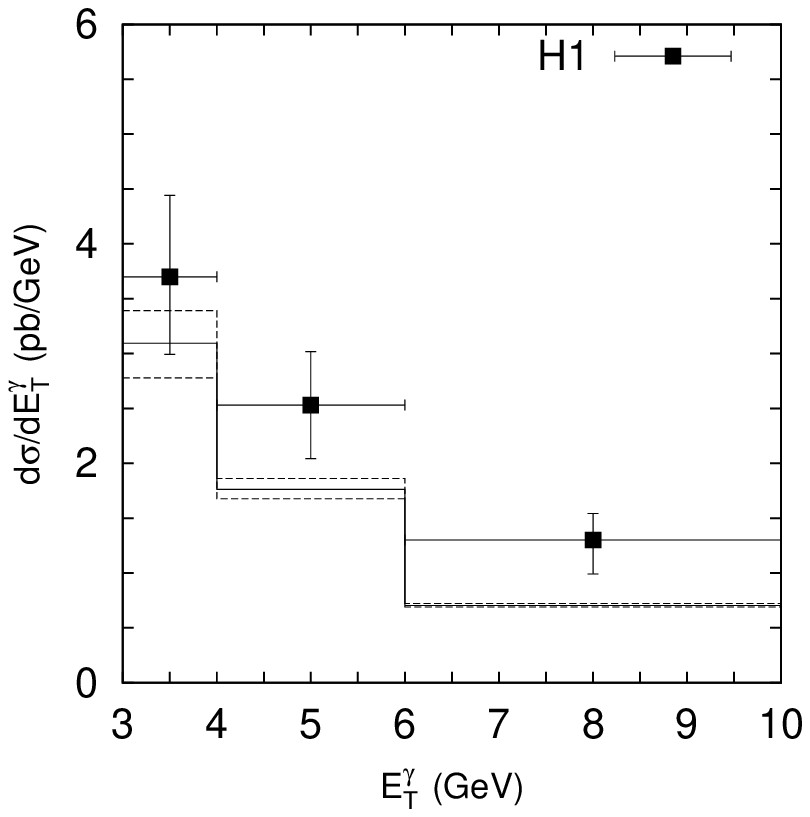, width = 8.1cm}
\epsfig{figure=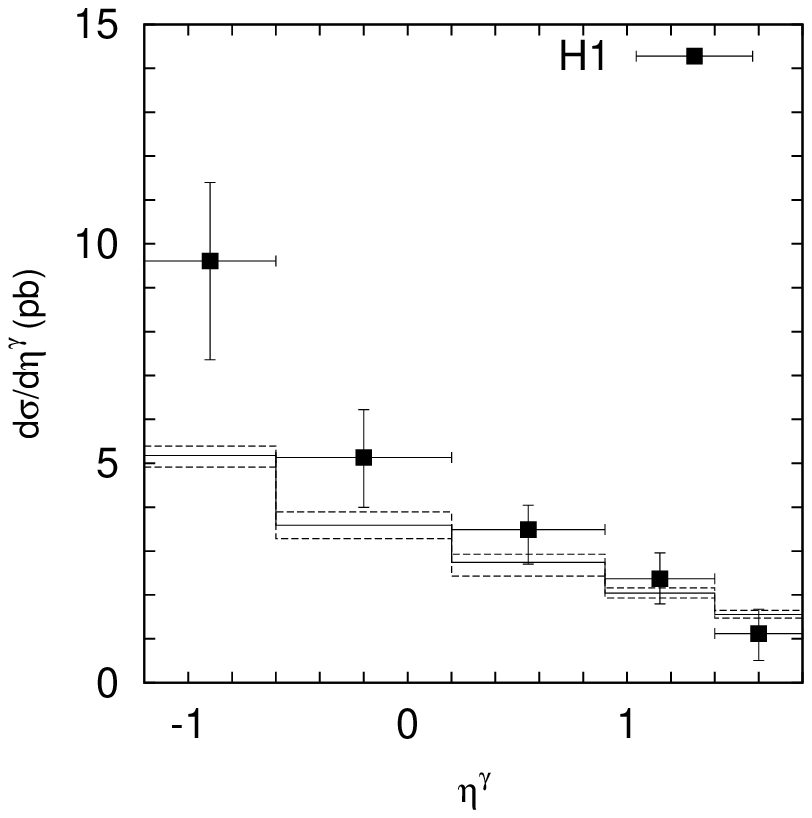, width = 8.1cm}
\caption{Differential cross sections of the deep inelastic
inclusive prompt photon production
as a function of $E_T^\gamma$ and $\eta^\gamma$
calculated at $3 < E_T^\gamma < 10$ GeV, $ - 1.2 < \eta^\gamma < 1.8$,
$40 < Q^2 < 150$~GeV$^2$, $E_e^\prime > 10$~GeV, $153^{\rm o} < \theta_e^\prime < 177^{\rm o}$, $y > 0.05$
and $W_X > 50$~GeV.
Notation of the histograms is as in Fig.~1. 
The experimental data are from H1~[3].}
\end{center}
\label{fig2}
\end{figure}

\begin{figure}
\begin{center}
\epsfig{figure=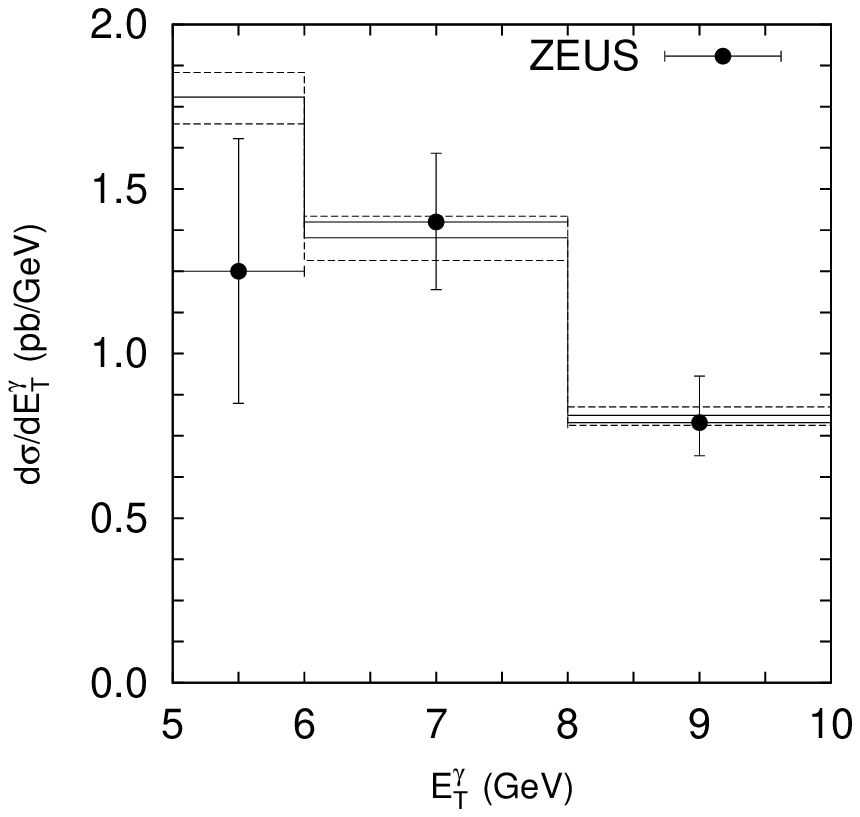, width = 8.1cm}
\epsfig{figure=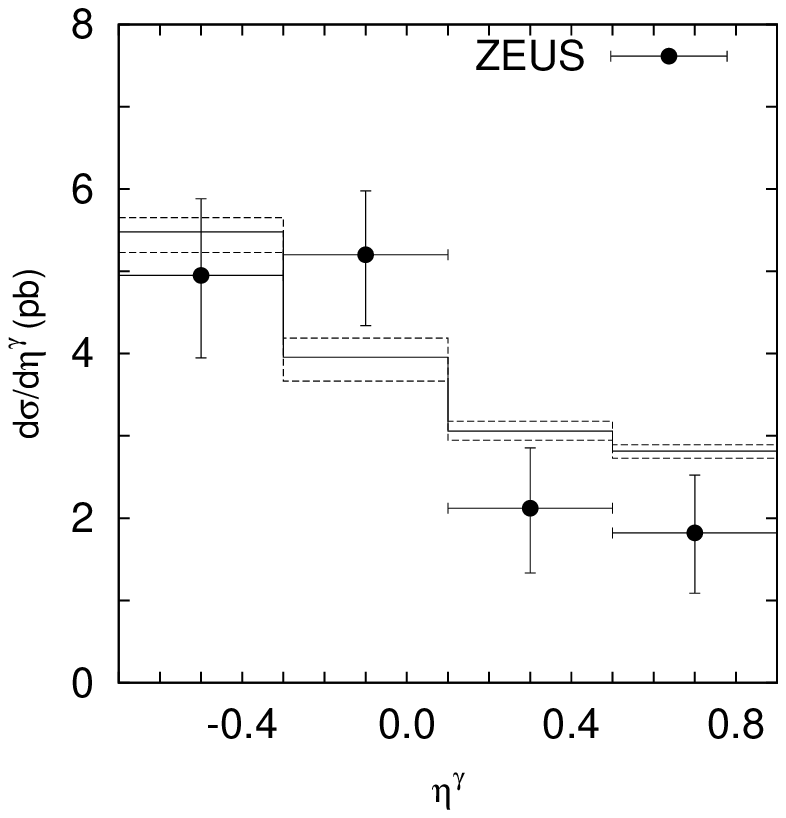, width = 8.1cm}
\caption{Differential cross sections of the deep inelastic
inclusive prompt photon production
as a function of $E_T^\gamma$ and $\eta^\gamma$
calculated at $5 < E_T^\gamma < 10$~GeV, $-0.7 < \eta^\gamma < 0.9$, $Q^2 > 35$~GeV$^2$,
$E_e^\prime > 10$~GeV and $139.8^{\rm o} < \theta_e^\prime < 171.8^{\rm o}$.
Notation of the histograms is as in Fig.~1. 
The experimental data are from ZEUS~[1].}
\end{center}
\label{fig3}
\end{figure}

\begin{figure}
\begin{center}
\epsfig{figure=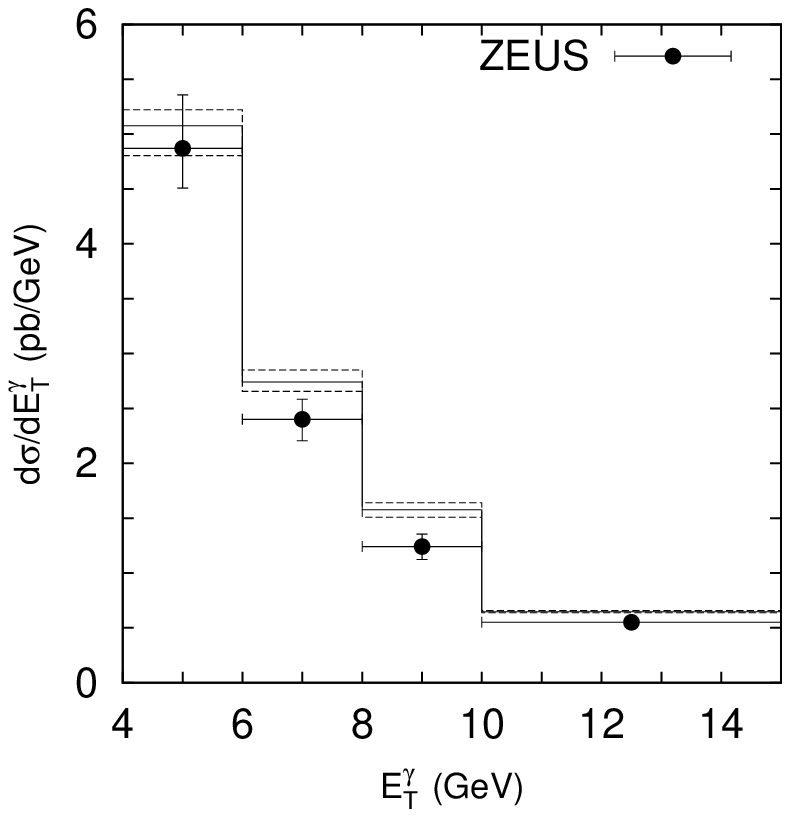, width = 8.1cm}
\epsfig{figure=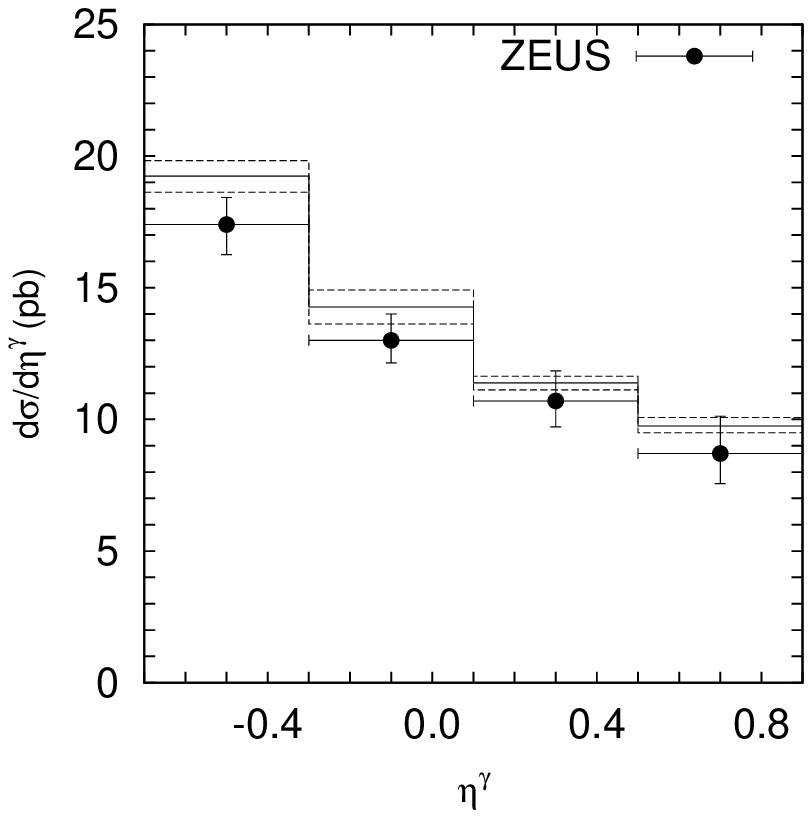, width = 8.1cm}
\epsfig{figure=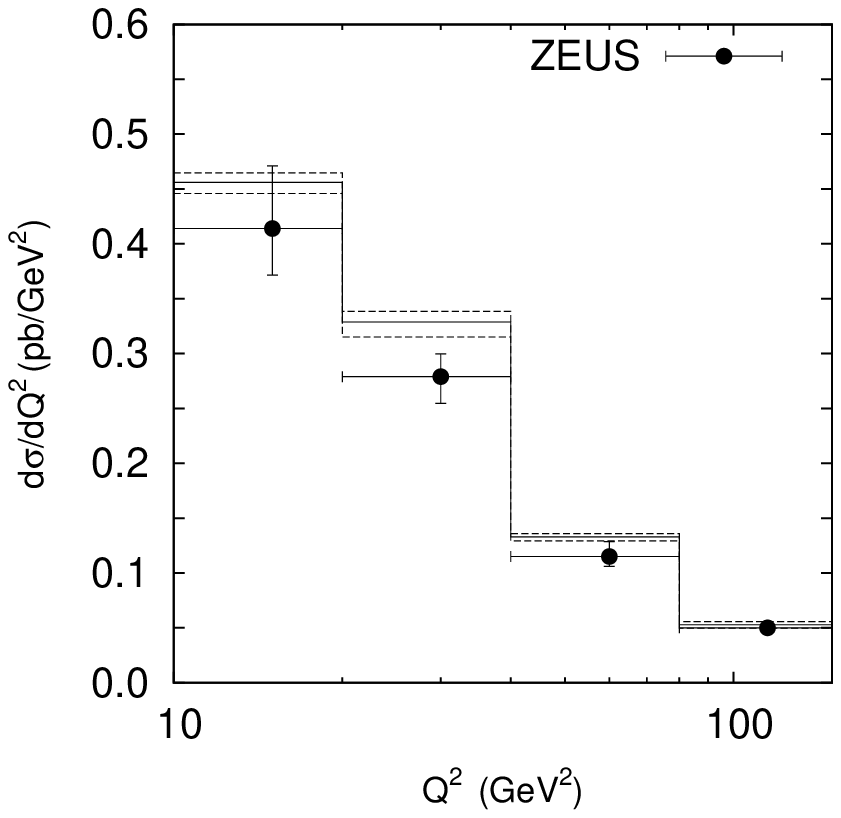, width = 8.1cm}
\epsfig{figure=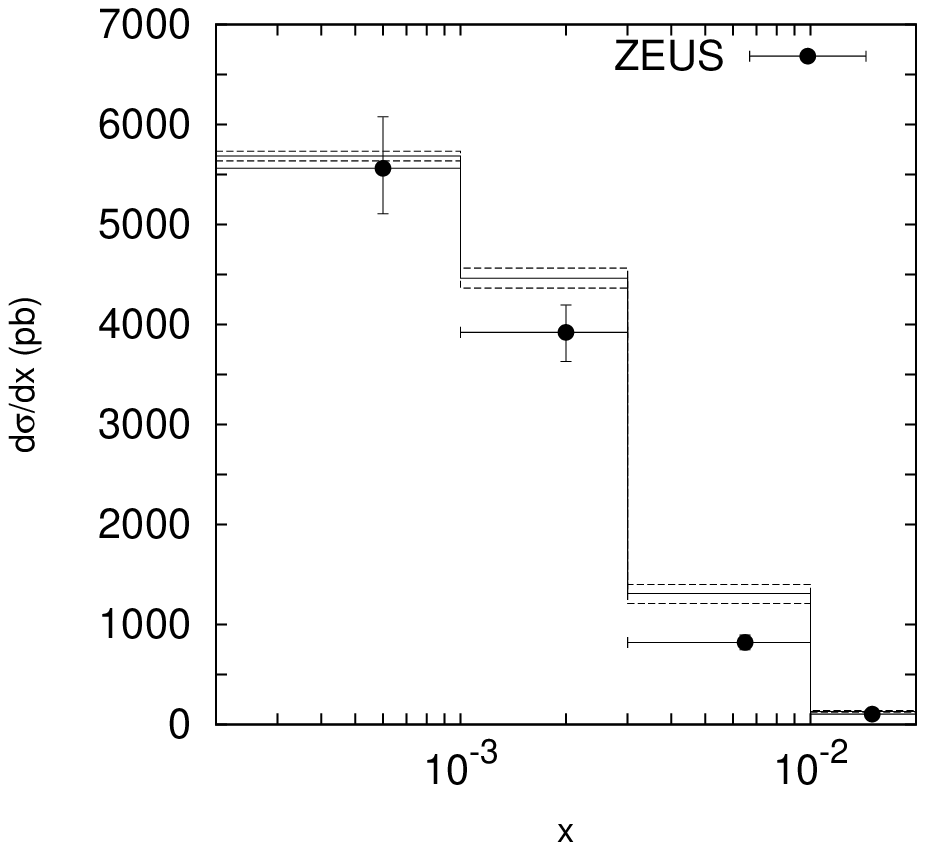, width = 8.1cm}
\caption{Differential cross sections of the 
inclusive deep inelastic prompt photon production at HERA
as a function of $E_T^\gamma$, $\eta^\gamma$, $Q^2$ and $x$
calculated at $4 < E_T^\gamma < 15$~GeV, $-0.7 < \eta^\gamma < 0.9$, $10 < Q^2 < 350$~GeV$^2$,
$E_e^\prime > 10$~GeV, $139.8^{\rm o} < \theta_e^\prime < 171.9^{\rm o}$ and $W_X > 5$~GeV.
Notation of the histograms is as in Fig.~1. 
The experimental data are from ZEUS~[2].}
\end{center}
\label{fig4}
\end{figure}

\begin{figure}
\begin{center}
\epsfig{figure=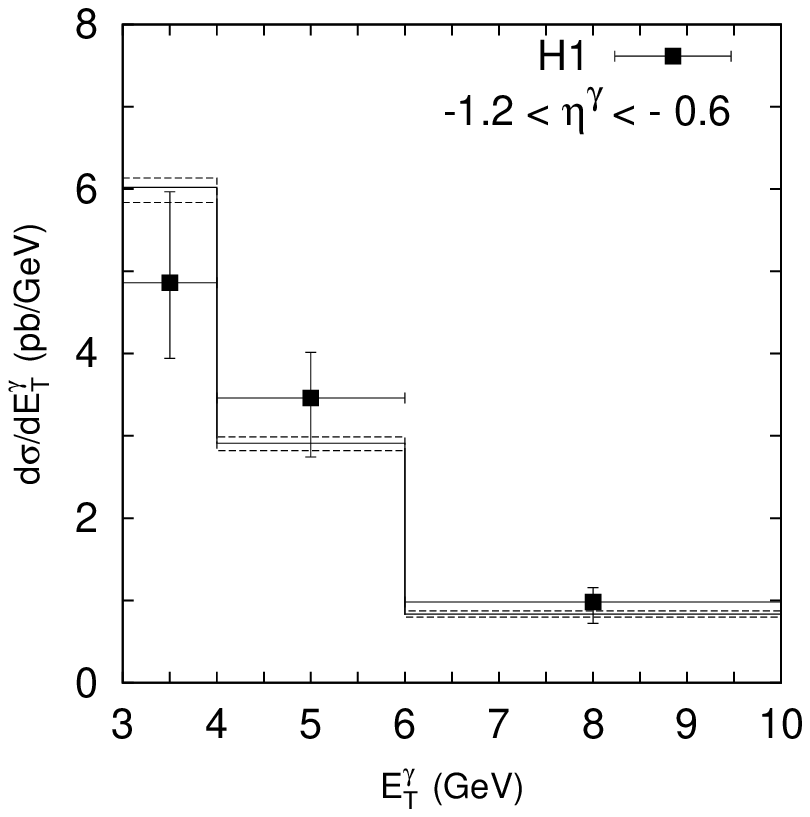, width = 8.1cm}
\epsfig{figure=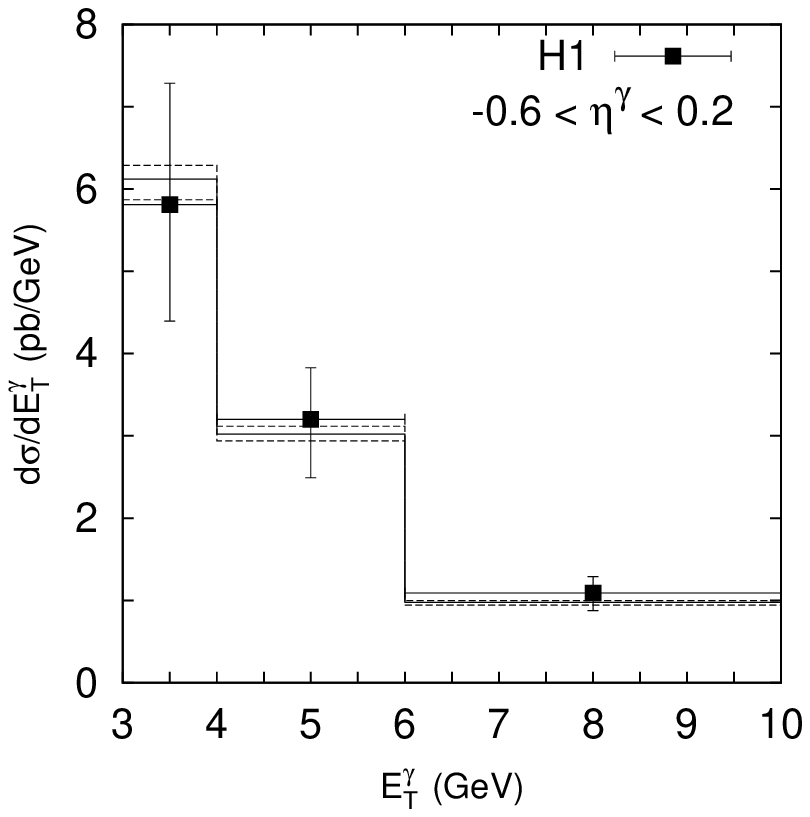, width = 8.1cm}
\epsfig{figure=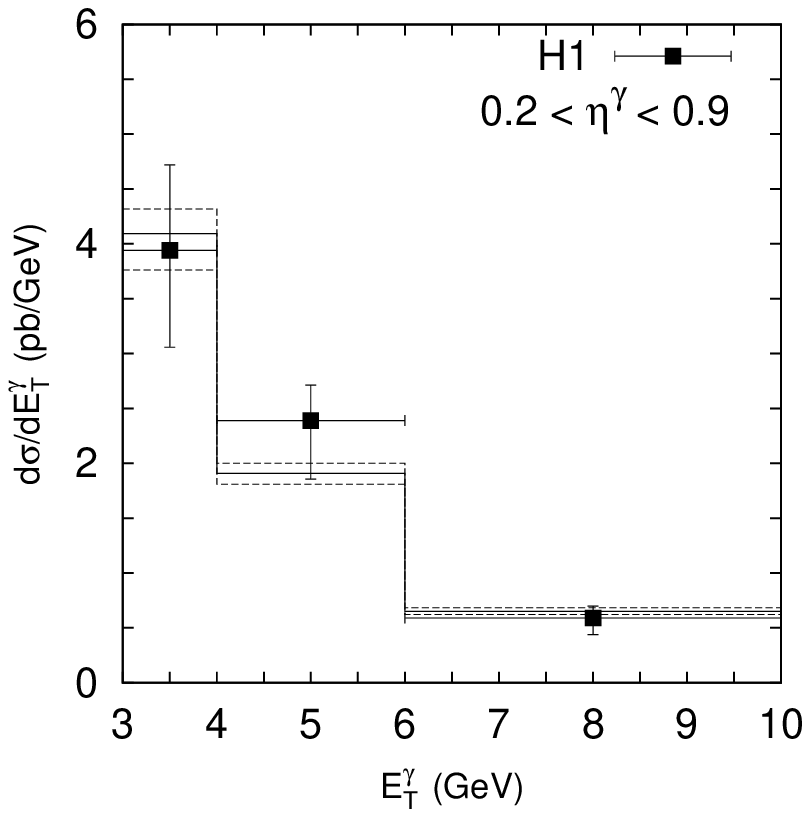, width = 8.1cm}
\epsfig{figure=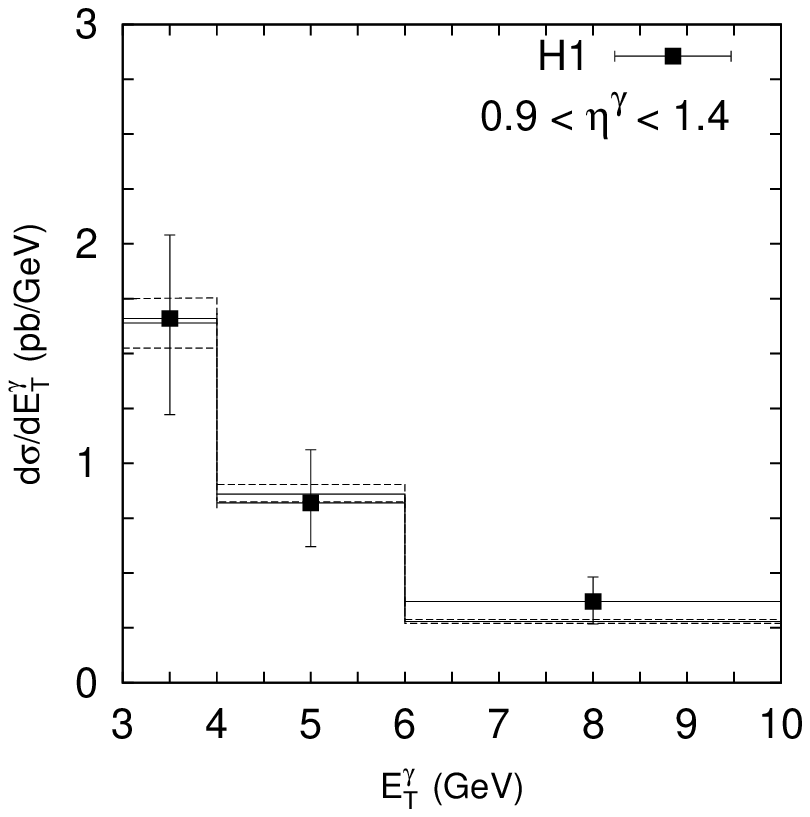, width = 8.1cm}
\epsfig{figure=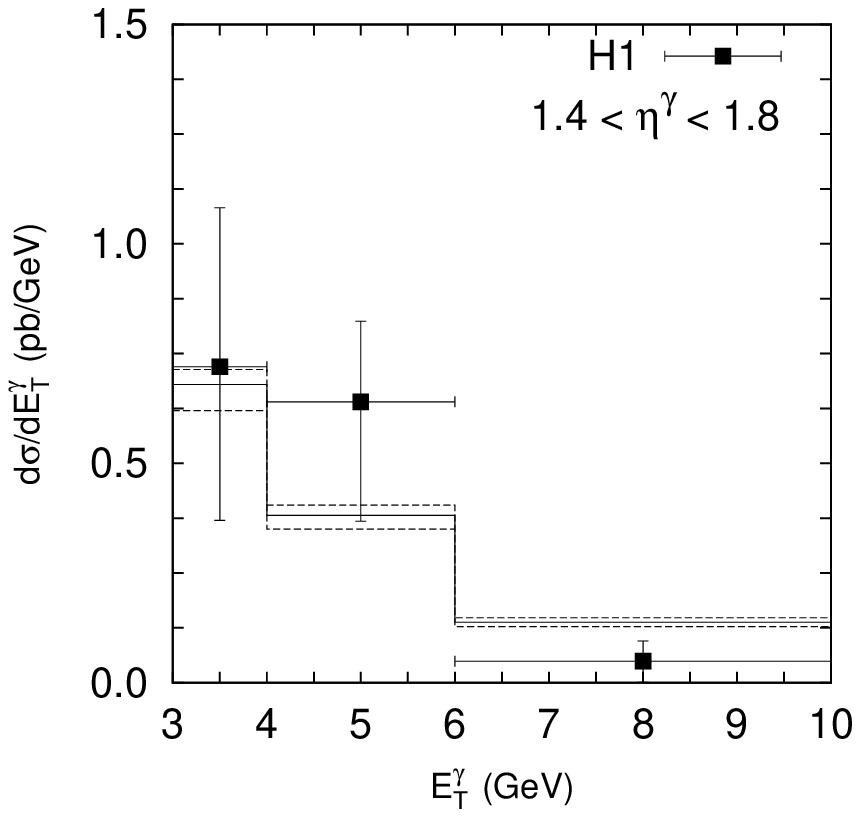, width = 8.1cm}
\caption{Differential cross sections of the deep inelastic
inclusive prompt photon production
as a function of $E_T^\gamma$ 
calculated for different subdivisons of the $\eta^\gamma$ range.
Notation of the histograms is as in Fig.~1. 
The experimental data are from H1~[3].}
\end{center}
\label{fig5}
\end{figure}

\begin{figure}
\begin{center}
\epsfig{figure=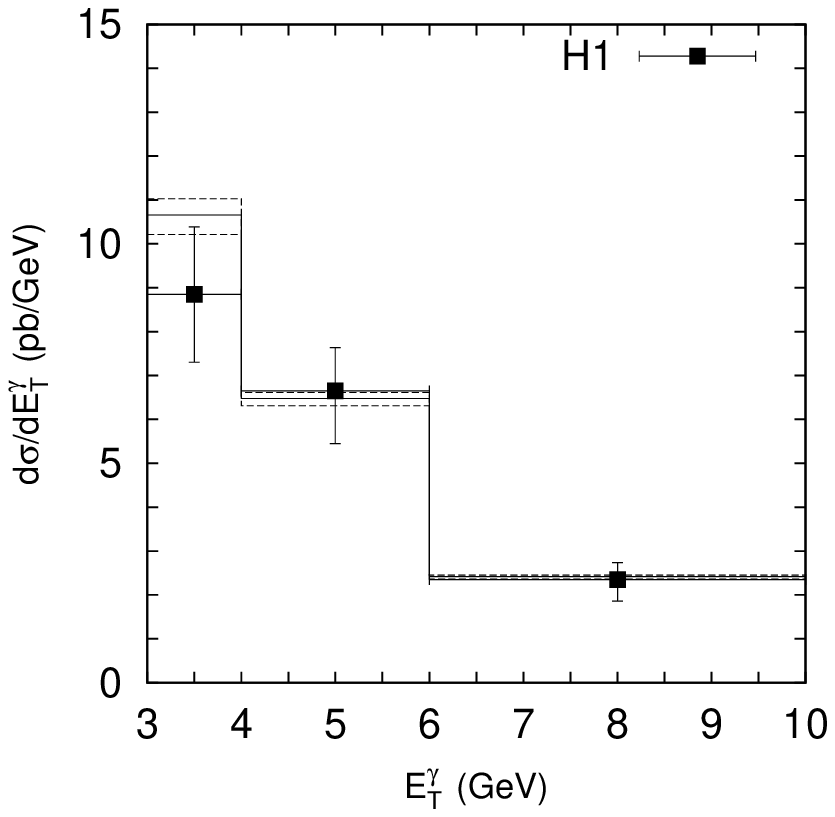, width = 8.1cm}
\epsfig{figure=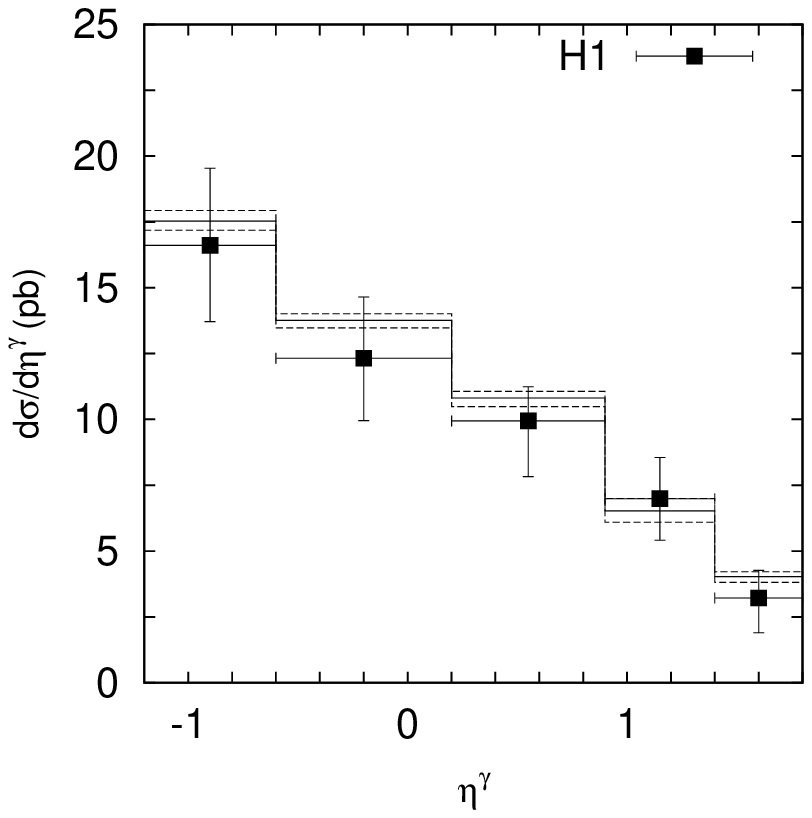, width = 8.1cm}
\epsfig{figure=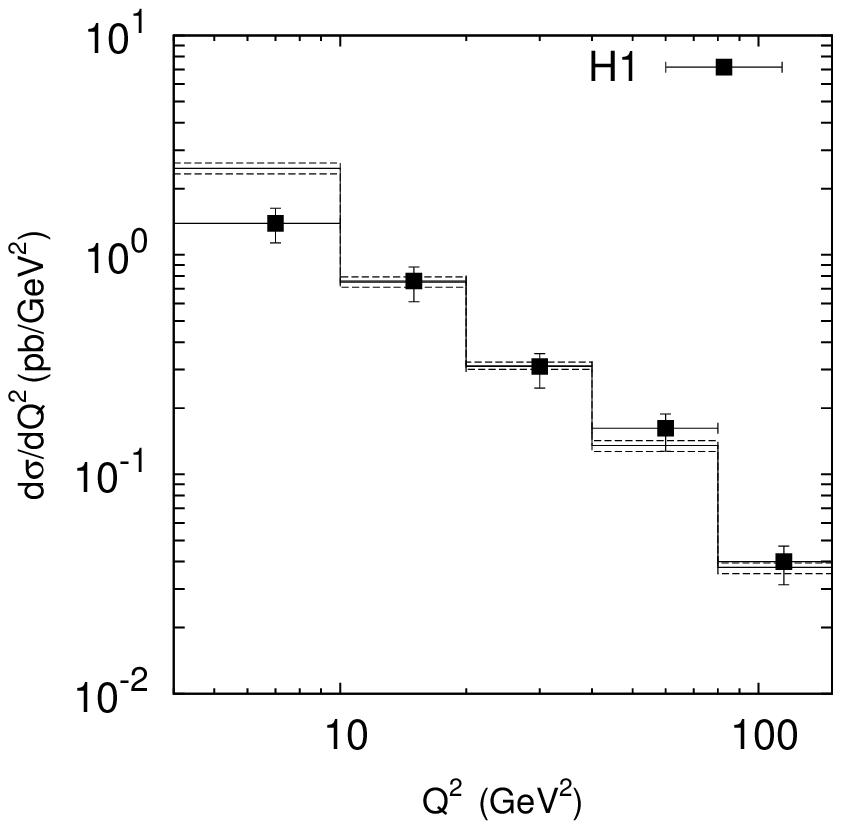, width = 8.1cm}
\caption{Differential cross sections of the 
deep inelastic prompt photon and jet associated production
as a function of $E_T^\gamma$, $\eta^\gamma$ and $Q^2$ calculated at
$3 < E_T^\gamma < 10$~GeV, $E_T^{\rm jet} > 2.5$~GeV, $ - 1.2 < \eta^\gamma < 1.8$,
$ - 1.0 < \eta^{\rm jet} < 2.1$, $4 < Q^2 < 150$~GeV$^2$, $E_e^\prime > 10$~GeV, $153^{\rm o} < \theta_e^\prime < 177^{\rm o}$,
$y > 0.05$ and $W_X > 50$~GeV.
Notation of the histograms is as in Fig.~1. 
The experimental data are from H1~[3].}
\end{center}
\label{fig6}
\end{figure}

\begin{figure}
\begin{center}
\epsfig{figure=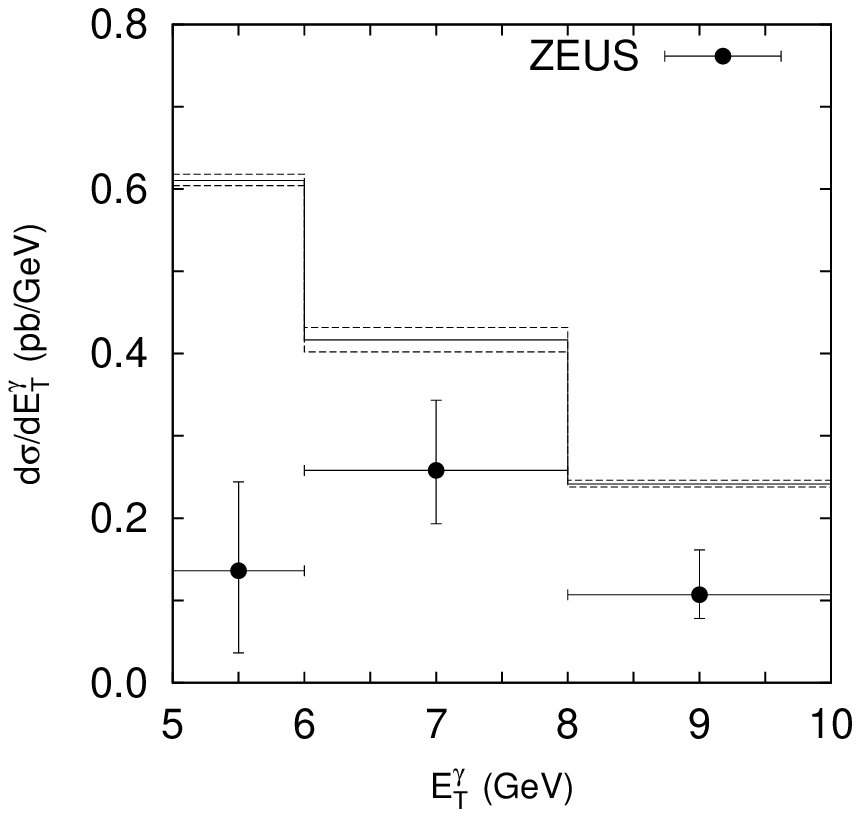, width = 8.1cm}
\epsfig{figure=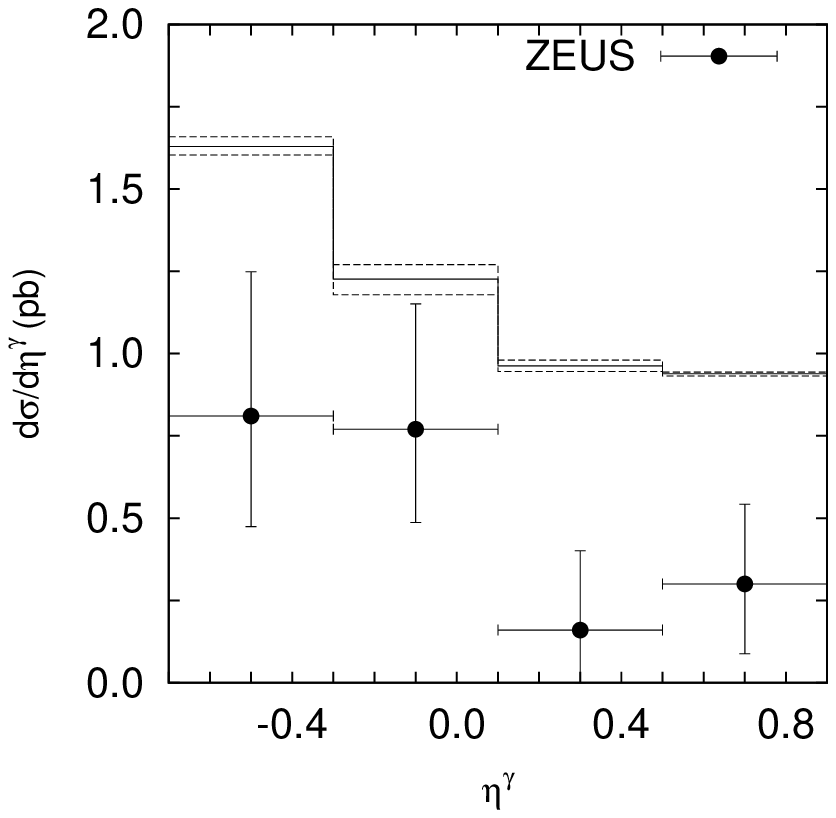, width = 8.1cm}
\epsfig{figure=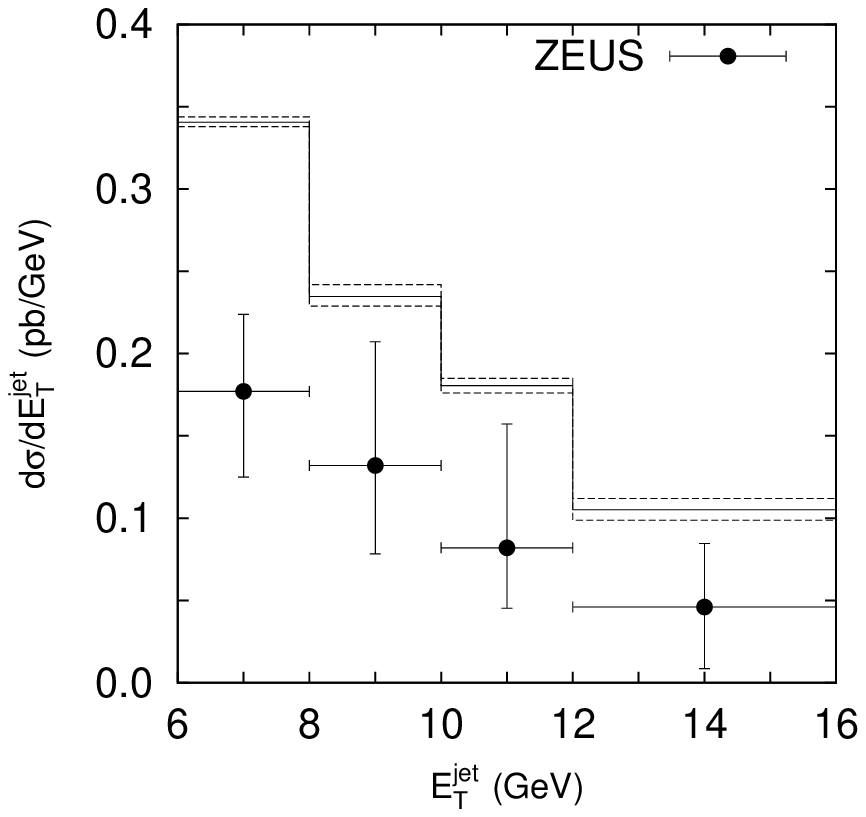, width = 8.1cm}
\epsfig{figure=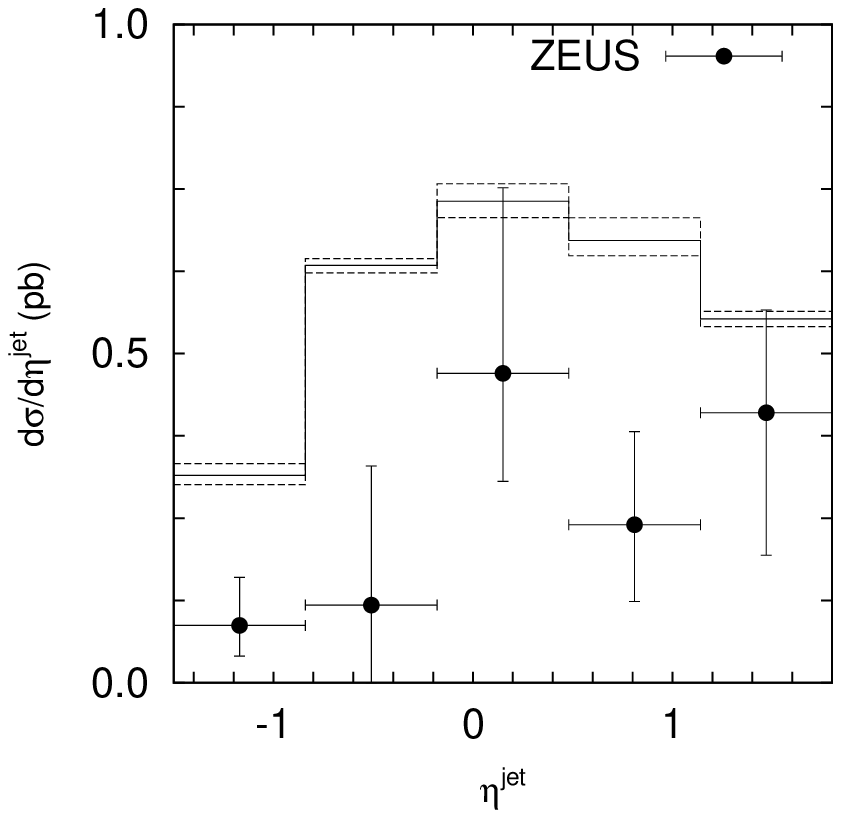, width = 8.1cm}
\caption{Differential cross sections of the 
deep inelastic prompt photon and jet associated production
as a function of $E_T^\gamma$, $\eta^\gamma$, $E_T^{\rm jet}$ and $\eta^{\rm jet}$ calculated at
$5 < E_T^\gamma < 10$ GeV, $E_T^{\rm jet} > 6$~GeV, 
$ - 0.7 < \eta^\gamma < 0.9$, $ - 1.5 < \eta^{\rm jet} < 1.8$,
$Q^2 > 35$~GeV$^2$, $E_e^\prime > 10$~GeV and $139.8^{\rm o} < \theta_e^\prime < 171.8^{\rm o}$.
Notation of the histograms is as in Fig.~1. 
The experimental data are from ZEUS~[1].}
\end{center}
\label{fig7}
\end{figure}

\end{document}